\documentclass[aps,prd,epsf,amsmath,amssymb,nofootinbib,11pt]{revtex4}
\usepackage{color,graphicx}
\usepackage{dcolumn}
\usepackage{bm}
\usepackage{mathrsfs}
\usepackage[T1]{fontenc}

\newcommand{\be}{\begin{equation}}
\newcommand{\ee}{\end{equation}}
\newcommand{\ba}{\begin{eqnarray}}
\newcommand{\ea}{\end{eqnarray}}
\newcommand{\ban}{\begin{eqnarray*}}
\newcommand{\ean}{\end{eqnarray*}}

\begin{document}

\title{Fluctuations in horizon-fluid lead to negative bulk viscosity
}
\author{Swastik Bhattacharya}\email{swastik@iisertvm.ac.in}

\author{S. Shankaranarayanan} 
\email{shanki@iisertvm.ac.in} 
\affiliation{School of Physics, Indian Institute of Science Education and Research Thiruvananthapuram (IISER-TVM), Trivandrum 695016, India}

\begin{abstract}
  Einstein equations projected on to a black hole horizon gives rise
  to Navier-Stokes equations. Horizon-fluids typically possess unusual features
  like negative bulk viscosity and it is not clear whether a statistical mechanical 
  description exists for such fluids. In this work,  we provide an explicit derivation 
  of the Bulk viscosity of the horizon-fluid based on the theory of fluctuations 
  {\it \`a la} 
  Kubo. The main advantage of our approach is that 
  our analysis remains for the 
  most part independent of the details of the underlying microscopic theory and hence the 
  conclusions reached here are model independent. We show that the coefficient
  of bulk viscosity for the horizon-fluid matches exactly with the value found from the
  equations of motion for the horizon-fluid.
 \end{abstract}

\maketitle

\section{Introduction}
In 1970's, it was found that the dynamics of Black-holes is formally
analogous to thermodynamics and is now widely viewed as the 
laws of black-hole thermodynamics \cite{BH-Thermo,Hawking-1975}.  
This offered support for a proposal made earlier by Sakharov
that Gravity may be an emergent phenomenon \cite{Sakharov,Padmanabhanreview,Jacobson,Parentani}. Around the same time,  
it was shown that the horizon of a black hole behaves like a viscous fluid i. e., 
horizon obeys Navier-Stokes equation \cite{Damour,Membrane} . Over the last few years, 
a growing body of evidence suggests that gravitational dynamics near the black-hole horizon, 
is analogous to the dynamics of a viscous fluid\cite{Shiraj,Paddy,Strominger}.

The current status of the fluid-gravity correspondence appears
somewhat similar to the laws of black-hole mechanics in early
70's. There seems to be a lot of similarities between the equations of
Gravity on the horizon and fluid mechanics. However, it is unclear
whether the mathematical similarities of the two sets of equations
describing two completely different systems suggest something more
physical. In particular, the question that needs to be answered is
whether the fluid-gravity correspondence can provide a
statistical-mechanical description of the degrees of freedom (DOF) that 
leads to black-hole entropy for a general black-hole spacetime. 

Such a program runs into difficulties primarily because the horizon-fluid does not behave like a
normal fluid.  In particular, it has a negative bulk viscosity ($\zeta$) for asymptotically flat space-times 
like Schwarschild, Kerr~\cite{Damour}. For asymptotically AdS black-holes,  for which the horizon-fluid 
has positive bulk viscosity,  there seems to exist such a description via a CFT on the 
boundary~\cite{Son,Shiraj}. However, it is not possible to provide a statistical mechanical description for 
the horizon-fluid corresponding to asymptotically flat space-times. For asymptotically flat 
space-times, we do not yet have any powerful physical principle that
can guide us from a microscopic description of Gravity to a fluid description. To be more specific,
there does not exist a microscopic theory dual (that exists for asymptotically AdS space-times) 
to the theory of Gravity for such space-times\footnote{ See \cite{Hartong} for an interesting attempt in 
this direction.}.

In view of this, we shall adopt a different approach. We assume that the horizon-fluid
undergoes statistical fluctuations and construct a statistical mechanical description 
of the fluid based on the theory of fluctuations~\cite{Kubo,Kadanoff,Kadanoff2}. We show that 
this approach provides mathematical tools to address some of the open problems in 
Fluid-Gravity correspondence. The dynamical 
phenomena that can be described by the theory of fluctuations mainly constitute a class of processes where
the system is typically slightly away from equilibrium and are broadly
known as Transport phenomena. In particular, the diffusion of heat,
electrical conductivity, and the shear and the bulk viscosity of the
fluid fall under this category.

In this approach, we take the view that the horizon-fluid possesses some kind of physical reality
beyond the formal similarity.  One advantage of this strategy is that our analysis would remain for
the most part independent of the details of the underlying microscopic
theory and hence the conclusions reached here would be model
independent. It also follows that in the absence of any microscopic theory, our approach here 
can only be phenomenological.  The whole analysis is performed within the
Mean Field Theory model of the horizon-fluid~\cite{MFT,GravEssay}. However the
main role of Mean Field Theory here is to motivate a choice of a
macroscopic variable in terms of which, one can write down the
Thermodynamic Potential for the fluid system. In some other model,
some other macroscopic variable might have to be chosen.  As will 
be clear, this does not change the physical analysis.

In what follows, we shall focus on the transport phenomena related
to the bulk viscosity of the fluid.  This is due to two
reasons. First, for bulk viscosity, we need to consider only a
homogeneous fluid system for the most part. This allows us to ignore
the effect of shear and simplifies the analysis to a great
extent. Second and more importantly, the Coefficient of bulk viscosity
is negative for the horizon-fluid and the reason for this is not
well-understood at present; it constitutes a challenge for any
statistical mechanical description of the horizon-fluid to compute
this quantity. Here we explicitly compute it using the theory of 
fluctuations and taking into account the teleological nature 
of the Black Hole event horizon~\cite{Membrane,antCausTranspo}.

The paper is organized as follows. In section II, we provide a broad 
outline of our approach to construct a theory of fluctuations for the
horizon-fluid. In section III, we discuss the Mean Field Theory model of the 
horizon-fluid and briefly review the earlier results that are essential to 
compute the transport coefficients. Section IV is the main part of the work 
where  we explicitly compute the Coefficient of bulk viscosity for the horizon-fluid. 
Finally, in section V, we discuss the implication of our results.

\section{Theory of fluctuations for the horizon-fluid: A broad outline}

In this section, we give a broad outline of the approach we follow to study 
the fluctuations and transport phenomena for the horizon-fluid. The four main 
ingredients that go in are as follows:
\begin{enumerate}
\item The macroscopic properties of the horizon-fluid must satisfy the
  black-hole constraints. For a Schwarzschild black-hole, the fluid
  must satisfy relations between total energy($E$), pressure($P$),
  temperature($T$), horizon area($A$) and the number of d.o.f.($N$).
 
\item The macroscopic properties of the horizon-fluid are specified by
  the collective properties of its microscopic constituents. The microscopic 
  constituents result in  statistical fluctuations of the macroscopic
  quantities. Such fluctuations will influence the evolution of a system from a non-equilibrium 
  state to a state in equilibrium.
 
\item  Any dynamical process that the event horizon of a black-hole
  undergoes corresponds to the horizon-fluid system moving from an
  initial non-equilibrium state towards an equilibrium state. However,
  so long as the evolution of the event horizon follows from a
  classical theory of Gravity, which is what we shall look at here,
  the system is not far from equilibrium in the corresponding fluid
  picture.
     
  The above assumption imposes a restriction on the kind of dynamical
  processes we consider on the Gravity side. In the fluid side,
  we are looking at processes that increase the entropy of the
  system. However, on the Gravity side, it is the area of the event
  horizon that is proportional to the entropy\cite{BH-Thermo,
    Hawking-1975, Wald} and even in a classical theory, it does not
  always increase. We do not take into account here the semiclassical
  effects like the decrease of black-hole area due to Hawking
  radiation \cite{Hawking-1975,Wald} as those would be quite small for
  large black-holes. In the classical theory, in order for the area of
  the event horizon to increase, some additional conditions have to be
  imposed\cite{Membrane}, like the condition that the black-hole is
  strongly asymptotically predictable \cite{Wald, Waldbk}.
 
\item The statistical fluctuations of the system are assumed to obey
  the Principle of Equipartition and Maxwell-Boltzmann Statistics.  
\end{enumerate}

For the fluid description, the relevant physical quantities vary 
slowly in space and time, a situation we will consider here. The
fluid-dynamical equations contain parameters, whose values cannot be
determined from fluid mechanics. These arise due to the changes in the
macroscopic variables of the system as a response to the external
influences that are not large. These parameters fall into two
classes: (i) Thermodynamic derivatives which relate the changes in
local Thermodynamic variables and (ii) transport coefficients like
viscosity and thermal conductivity that relate the fluxes of
thermodynamic quantities to the gradients of the local variables. These 
parameters can only be obtained from a more fundamental theory\footnote{See 
Ref.~\cite{Kadanoff} for a good discussion on this topic. Our treatment of this
  issue is very much in the same spirit.}. The theory of fluctuations
that we construct here for the horizon-fluid works at this level.

Statistical fluctuations of normal fluids is well-understood~\cite{Onsager}. 
The horizon-fluid, on the other hand, is an unusual fluid. Unlike normal fluids,
it is a one parameter system, whose energy, pressure, temperature,
volume(area in this case) are not independent of each other. It also
has negative bulk viscosity. Here we explicitly show that it is
possible to construct a theory of the Fluctuations in the Horizon
Fluid in analogy with normal fluids. As will be seen, our approach is
quite general and does not depend on the particular Mean Field Theory
model that we use to compute the Coefficient of
bulk viscosity ($\zeta$).

The basic strategy to construct the theory and compute transport
coefficients is as follows:
\begin{enumerate}
\item The horizon-fluid for a Schwarzschild black-hole is a one
  parameter system\cite{MFT,Skakala}. We identify a relevant
  macroscopic variable and write down the Thermodynamic Potential in
  terms of it. The equilibrium state corresponds to a minimum of the
  Thermodynamic Potential.
 
\item The fluctuations about the equilibrium position of the system
  are about this minimum value. This implies that the leading
  order variation in the potential would be of second order in that variable.
 
\item We write down a Langevin equation that governs the transport
  processes to be considered.
 
\item We make the assumption that the random fluctuations that the
  system undergoes are of a much shorter time-scale than the time
  period over which the system goes over from one state to another due
  to external influence. These latter processes can also be viewed as
  fluctuations that the system undergoes; only they are of a much
  longer time-scale. This makes the probability distributions for the
  random short-time fluctuations and the long time ones independent of
  each other. It is possible to compute the correlation functions that
  contain all the information about the system from the assumptions that the
  random fluctuations obey the principle of Equipartition and both the
  probability distributions are given by Maxwell-Boltzmann.  Another
  necessary input is the black-hole constraints, which reflects the
  fact that Horizon fluid is actually a one parameter system.
 
\item The Transport coefficients for the horizon-fluid can then be
  computed using a method given by Kubo\cite{Kubo} either in real
  space or in frequency space.  
 \end{enumerate}

\section{Mean Field theory and Fluctuations about the Equilibrium}

In this section, we provide key results of Refs. \cite{GravEssay} that forms the basis 
for computing the transport coefficients.  In particular, we shall sketch the Mean Field Theory
for the horizon-fluid for a Schwarzschild black-hole. After this, we
shall set up a theory of the statistical fluctuations within a Mean
Field Theory setup. In the first subsection, we list all the useful relations of
black-hole thermodynamics.

\subsection{Constraints from black-hole Physics}

The Einstein equations projected on the event horizon of a
Schwarschild black hole can be described by a $(2+1)-$dimensional
fluid that resides on the event horizon of the black hole \cite{Damour}.  
In particular, in Damour's approach, the Navier-Stokes equation\footnote{Actually the equation found by 
Damour is not exactly same as the Navier-Stokes equation but very similar to it.}
governing the horizon-fluid dynamics is given by: 
 \begin{equation}
   \frac{D\Pi_A}{dt}= 
-\frac{\partial}{\partial x^A}(\frac{\kappa}{8\pi}) 
+2 \frac{1}{16\pi}\sigma^B_{A \textbar B}
-\frac{1}{16\pi}\frac{\partial\theta}{\partial x^A} - l^aT_{aA}, \label{NavStok}
 \end{equation}
 where, $\mathbf{\sigma}$ is the shear tensor for the null congruence
 on the horizon, $\mathbf{l}$ is the vector normal to the horizon and
 $x^A$ denotes the coordinates on the null hypersurface perpendicular
 to the null generators of the horizon. The vertical bar denotes the
 covariant derivative with respect to the induced metric on the null
 hypersurface. $\mathbf{\Pi}$ can be thought of as the momentum
 density of the horizon-fluid. The coefficient of bulk ($\zeta$) and 
 Shear viscosity ($\eta$), respectively,  are given by  %
 \begin{equation}
  \zeta=-\frac{1}{16\pi}; \space \eta= \frac{1}{16\pi}. 
 \end{equation}
The volume of the fluid is the area of the horizon ($A$).
 The temperature of the horizon ($T$) is the temperature of the fluid and 
 the total energy of the fluid is given by the Komar mass of the black hole \cite{Skakala}. The
equation characterizing the fluid is given by \cite{Skakala},
\begin{equation}
 P= \frac{T}{4}=\frac{E}{2A}. \label{eqs}
\end{equation}
The parameter space of the Schwarzschild black-hole is one
dimensional. So $E$, $T$, $A$ are not independent, but must obey the
two constraints: $E= E(A)$ and $E=E(T)$. The constraint equation
between $E$ and $A$ is given by\cite{Skakala},
\begin{equation}
 A = 16\pi E^2. \label{AE}
\end{equation}
The constraint equation relating the black hole mass and the Hawking
temperature is
\begin{equation}
 E= \frac{1}{8\pi T}. \label{ET}
\end{equation}

\subsection{The Mean Field Theory model for the horizon-fluid}
We start by assuming that the Horizon-fluid forms a condensate at a
critical temperature. The justification comes from two lines of
arguments. First, the evidence provided by Carlip\cite{Carlip} that
the black-hole horizon has some properties that exhibit
universality. This indicates that the physics near the horizon is that
of a system near a critical point.  Second, recently, Skakala and
Shankaranarayanan\cite{Skakala} modelled the fluid as a Bose gas with
$N$ particles and found that all the particles stayed in the ground
state for large horizon radius. This suggests that horizon-fluid forms
a BEC at some critical temperature $T_c$. The conditions underlying
this are\cite{MFT}:
\begin{enumerate}
\item There is a temperature $T_c$ (critical temperature), at which,
  all the $N$ fluid degrees of freedom on the horizon form a
  condensate.
\item The system remains close to the critical point.
\end{enumerate}
An immediate consequence of these assumptions is the deduction of the
relation between $N$ and $A$.  Since the system forms a condensate at
$T_c$, nearly all the microscopic d.o.f.s would be in the ground state
near the critical point. As there is only one scale in the problem,
the total energy of the system can be expressed in the form,
\begin{equation}
 E= N \alpha T, \label{NE}
\end{equation}
where, $\alpha$ is a constant. Using the other black-hole constraints,
one can write a relation between $N$ and $A$,
\begin{equation}
 N= \frac{A}{2\alpha}, \label{NA}
\end{equation}

One can describe this critical system, a homogeneous fluid, using Mean
Field Theory.  Taking the order parameter to be
\begin{equation}
 \eta= \sqrt{kN}. \label{eta}
\end{equation}
The thermodynamic Potential $\Phi$ is %
\begin{equation}
 \Phi= \Phi_0+a(P)(T-T_c)\eta^2+B(P)\eta^4. \label{phiLT}
\end{equation}
where $k$ is a constant, coefficients $a$ and $B$ are determined by
the relation $P=-TA$. One can determine the entropy of the fluid in
$\eta\ne0$ phase and show that it leads to the Bekenstein-Hawking
entropy, $S= \frac{A}{4}$ \cite{MFT}. The Horizon-fluid is near a
critical point and when the system goes over to the ordered phase, its
entropy is the same as the black-hole entropy.  This formalism can be
extended to include black-holes in AdS background\cite{MFT}. 

\subsection{Fluctuations about the Equilibrium Position}
What has been described so far, corresponds only to static situations
in the gravity theory. The dynamics of the black-hole event horizon
can be described by the equations of motion of a fluid. The basic idea
that underlies our Statistical Mechanical description of the Horizon
Fluid is that the dynamical evolution of the black-hole event horizon
in the fluid picture corresponds to the fluid system moving from a
state slightly away from the equilibrium towards a state in
equilibrium \cite{GravEssay}.

When one does not look at the dynamics of the evolution and only considers the initial and the 
final states, then the change in the energy and the entropy of the fluid in the process are seen to be 
related via the 
First Law of Thermodynamics. This corresponds to the Physical Process First Law of black-hole Thermodynamics
[See Appendix A]. However, if one considers how the dynamical evolution of the fluid takes place, then such a 
process can be described by a Langevin equation. It can then be shown that this equation is 
similar to the energy conservation equation of the Horizon Fluid or the Raychaudhury equation for 
the null congruences on the horizon[See Appendix A].

Let us now determine how the fluctuations from the equilibrium position can be expressed within the 
Mean field Theory.  The one parameter Schwarzschild Black hole system corresponds to the fluid system at 
equilibrium. Fluctuations, assumed here to be isothermal, do not obey all the constraint equations. As this 
is a one parameter system, the fluctuations can be 
charecterised by the deviations of the order parameter $\eta$ around its mean 
value at equilibrium. This implies that not all the constraint equations are satisfied 
when the system is away from equilibrium. In the Mean field Theory, the equilibrium state corresponds to any
 one of the minima of the double well for the Thermodynamic Potential
 $\Phi$ of the horizon-fluid system. This state is described by the
 value of the order parameter $\eta$,
\begin{equation}
 \eta_{min}^2= \frac{a(T_c-T)}{2B}. \label{etaEq}
\end{equation}
Let, $\delta\eta$($\delta N$) denote the change in the value of the
order parameter(number) due to fluctuations, then the change in the
potential is
\begin{equation}
\delta\Phi= \frac{1}{4}\alpha (T-T_c)\frac{\delta N^2}{N_0}. \label{deltaphi}
\end{equation}
This relation will be useful when we compute the coefficient of Bulk Viscosity.

 
 \section{bulk viscosity of the horizon-fluid}
 
 In this section, using the theory of fluctuations \cite{Kadanoff,Kubo}, we compute the 
 coefficient of bulk viscosity ($\zeta$) of the horizon-fluid.
 
 The transport coefficient for a particular process can be determined 
 by identifying the relevant current and computing the autocorrelation function of 
 the current~\cite{Kadanoff,Kubo,Onsager,Zwanzig,Dufty}. In the case of the horizon-fluid, 
 we need to identify the current corresponding to the change in the fluid volume  (area in this case). 
Unlike normal fluids, the horizon-fluid is a one parameter system. Hence, it
 is not completely straightforward to generalise the formula of the ordinary fluid to the case of the
 horizon-fluid. 
 
We proceed in the following manner: (i) Obtain the formula for $\zeta$ 
 for a normal fluid and relate the corresponding current to an entropic force acting on
 the system. (ii) Obtain $\zeta$ in terms of 
 an autocorrelation function of the entropic force for a normal fluid.  (iii) Extend the formula 
 to compute $\zeta$ for the one parameter horizon-fluid system for anti-causal transport processes.

\subsection{$\zeta$ for a normal fluid in Fluctuation Theory}

For a 3-dimensional Newtonian fluid, Kubo's method \cite{Kubo} can be used to
calculate $\zeta$. If the particles obey the Principle of Equipartition and follow 
Maxwell-Boltzmann statistics, then $\zeta$ is given by \cite{Kadanoff},
\begin{equation}
 \zeta= \left(\frac{1}{9}\right) \frac{1}{VKT}\int_0^\infty dt 
\sum\limits_a\sum\limits_b\langle J^{aa}(0)J^{bb}(t)\rangle \, ,
\label{BVformula}
\end{equation}
where, 
\begin{equation}
 J^{ab}=  \delta_{ab} \,  \delta (PV)=V \delta P \, \delta_{ab} \, .
 \label{JEq}
\end{equation}
$\delta$ denotes the variation in any quantity \cite{Onsager,Zwanzig} and the factor 
$\frac{1}{9}$ comes from the number of dimensions ($3-$dimensions). The 
range of integration is between $[0, \infty)$ as only causal processes are being considered. It is to be 
noted that the range of integration will be different for a horizon-fluid. 

Substituting (\ref{JEq}) in  \eqref{BVformula}, we get,
\begin{equation}
 \zeta=  \left(\frac{1}{9}\right) \frac{V}{KT} 
\sum\limits_a\sum\limits_b \delta_{aa}\delta_{bb}
\int_0^\infty dt\langle\delta P(0)\delta P(t)\rangle 
= \frac{V}{KT} \int_0^\infty dt\langle
\delta P(0)\delta P(t)\rangle. 
\label{zetaP}
\end{equation}
Keeping in mind that our primary concern for this work is the horizon-fluid --- which is 
highly constrained --- the change in the pressure of the fluid can then 
be related to the entropic force acting on the system. We show  
this~\cite{BVMD} by assuming an idealised situation, where the system is isotropic and resides inside a cube.
Let $ \delta V_{Strain}= \delta V_{Compress}/V$. Let us assume that the energy dissipated in 
this process is given by $\delta E_d$ and
\begin{equation}
 \delta E_d= -F_{Th} \delta V_{Strain}, \label{Diisip}
\end{equation}
where, $F_{Th}$ is the entropic force. If we assume that the work done
on the system is totally lost i. e.  it gets converted to heat, then due to the 
isotropy of the system, one can write,
\begin{equation}
 \delta E_d= -3 \delta P \delta V_{Compress}
= (3 \delta P V) \delta V_{Strain}. \label{PEd}
\end{equation}
Comparing \eqref{Diisip} with \eqref{PEd}, we identify,
\begin{equation}
 F_{Th}= 3\delta P V. \label{Fth}
\end{equation}
Recalling the expression for the current \eqref{JEq}, we identify 
entropic force $F_{Th}$ as
\begin{equation}
 F_{Th}^{ab}=J^{ab} = \delta P \, V \, \delta_{ab} \, . \label{JF}
\end{equation}

Substituting the above expression in \eqref{zetaP}, we get,
\begin{equation}
\zeta= \frac{1}{VKT}\int_0^\infty dt \langle 
F_{Th}(0)F_{Th}(t)\rangle. \label{zetaF1}
\end{equation}
This expression is also valid for the horizon-fluid with modified limits of integration.

\subsection{$\zeta$ for the horizon-fluid in Fluctuation Theory}

Evaluation of $\zeta$ for the horizon-fluid will be done in two steps: 
First, we shall write down the entropic force acting on the horizon-fluid, 
when the system is not in equilibrium.  Then we have to evaluate an
autocorrelation function for the entropic force, which requires a crucial input about the horizon 
properties. 

\subsubsection{Entropic force for the horizon-fluid}

As discussed earlier, we associate the deviation from the equilibrium 
to the fluctuation around one of the minima of the thermodynamic potential,
$\Phi$ of the horizon-fluid. The change in the potential ($\delta\Phi$) is given by
\eqref{deltaphi}. The energy lost in this process is the heat ($\delta Q$) generated 
during the process. $\delta Q$ can be determined from the First Law of Thermodynamics,
$\delta Q= T \delta S$. From \eqref{deltaphi}, the
change in entropy in this process comes out to be
\begin{equation}
 \delta S= -\frac{1}{8} \frac{\delta A^2}{A}. \label{deltaS}
\end{equation}
The negative sign of the change in
entropy denotes that the system is out of equilibrium. Since $\delta
E_d= \delta Q$, we have 
\begin{equation}
 \delta E_d= \frac{T}{8} \frac{\delta A^2}{A}
= \frac{T}{8} (\frac{\delta A}{A})^2A. \label{Ed1}
\end{equation}

From \eqref{Ed1}, we get, 
\begin{equation}
  d(\delta E_d)= \frac{T}{4} \frac{\delta A}{A}d(\delta A)
 \label{Ed2}
\end{equation}
The above expression is of the form 
\begin{equation}
 d(\delta E_d)= F_{Th} d(\delta A_{Strain}). \label{FHF}
\end{equation}
Comparing \eqref{Ed2} and \eqref{FHF}, gives 
\begin{equation}
 F_{Th}= P\delta A. \label{EntrForce}
\end{equation}

\subsubsection{The Teleological nature of event horizon and Anti-causal
  Transport Process in the horizon-fluid}
 
As mentioned earlier, only causal response is seen in normal-fluids in the presence of external influence. 
However, the same is not true for the horizon fluid. To understand why this happens, let us first 
look at the evolution of the black-hole event horizon.  
For the event horizon, the response to any external influence is
anti-causal. In particular, if matter- energy falls through the event horizon, 
then the area of the event horizon increases till the matter-energy passes through the horizon. 
This is not unphysical as the event horizon of a black hole 
is defined globally in the presence of the future lightlike infinity~\cite{Membrane}

Due to this unusual property of the horizon, the horizon-fluid
also exhibits anti-causal response, i.e. the response of the horizon
takes place before the external influence occurs\cite{Membrane}. Since
from the fluid point of view, the system is initially out of
equilibrium and slowly moves towards the final state in equilibrium,
it follows that the external influence brings the system to
equilibrium, so that there is no further evolution of the system from the state it is in.  This is referred 
to as the teleological nature of horizon~\cite{Membrane}.  

For a class of systems, it has been shown in the literature that if 
the system exhibits anti-causal transport process, then the anti
causal transport coefficients have an opposite sign to their causal
counterparts\cite{antCausTranspo}.  
For normal fluids, external influence drives the system out of
  equilibrium. For the horizon-fluid, it is the reverse; the system moves 
  towards equilibrium in anticipation of the external influence like infusion of energy into
  the fluid. This is the anti-causal response of the horizon-fluid.

The anti-causal response can be incorporated by defining
$\delta A_a(t)$ as $\delta A_a(t)= \delta
A(t)\theta(-t)$ and the anti-causal entropic force ($F_{Th}^{(a)}$) is then given by,
\begin{equation}
 F_{Th}^{(a)}(t)= PA_0\delta A_a(t)= P A_0 \delta A(t)\theta(-t). \label{Fac}
\end{equation}
One can also arrive at the same relation using Green's function to
calculate $\zeta$.  When determining $\zeta$, we have to determine the
autocorrelation function of $F_{Th}^{(a)}(t)$.

\subsubsection{Computation of $\zeta$ using Kubo's method}

In the rest of this section, we compute $\zeta$ in the real space. In Appendix B, we 
compute the same in the Frequency space. It is important to note that both the 
computations lead to the same value of $\zeta$.  $\zeta$ is given by, 
\begin{equation}
 \zeta= \frac{1}{AKT}\int_{-\infty}^\infty dt \langle F_{Th}^{(a)}(t)F_{Th}^{(a)}(0)\rangle \, . \label{zetaFa}
\end{equation}
Substituting for  $F_{Th}^{(a)}$ from \eqref{Fac}, we get
\begin{equation}
 \zeta= \frac{P^2}{AKT} \int_{-\infty}^\infty dt \langle \delta A(t)\delta A(0)\rangle \theta(-t). \label{zetaAR}
\end{equation}
Using \eqref{eqs} and \eqref{NA}, \eqref{zetaAR} can be written as 
\begin{equation}
 \zeta= \frac{\alpha^2T}{4AK}\int_{-\infty}^\infty dt \langle \delta N(t)\delta N(0)\rangle . \label{zetaNR}
\end{equation}
To proceed further, one needs to know the functional form of $\delta
N(t)$. Since our goal here is to provide an analytical expression for the transport 
coefficients of the fluid, which corresponds to the long wavelength limit 
of the microscopic theory \cite{Onsager,Kadanoff, Dufty}, the long wavelength 
(small frequency) limit of the above expression leads to 
\begin{equation}
  \zeta=\lim_{\epsilon\rightarrow 0}Im[\frac{\alpha^2T}{4AK} \langle\delta N^2(0)\rangle \int_{-\infty}^\infty dt \exp{i(\omega-i\epsilon)t}\theta(-t)]. \label{zetaNR1}
\end{equation}
Hence,%
\begin{equation}
  \zeta= -\frac{\alpha^2T}{4AK}\frac{\langle\delta N^2(0)\rangle}{\omega}. \label{zetaNR3}
\end{equation}
%
To obtain $\zeta$, we  need to determine $\langle \delta N^2(0) \rangle$ and $\omega$.  
Note that $\omega$ corresponds to the lowest possible frequency of the system, which for the 
horizon-fluid is related to the horizon cross-section. For Schwarschild, we have 
\begin{equation}
 r_h= 2 \, E
\end{equation}
Using \eqref{ET}, we can write this as 
\begin{equation}
 r_h=  \frac{1}{4\pi T}.\label{rhT}
\end{equation}
The circumference, $s$ of the black-hole cross-section is given by $s=
2\pi r_h=\frac{1}{2T}$. Then the wavenumber $k$ of the wave modes is
\begin{equation}
 k=\frac{2\pi}{\lambda}= 4\pi T. \label{kT}
\end{equation}
Hence, 
\begin{equation}
 \omega= k= 4\pi T.\label{omega}
\end{equation}

To determine $\langle\delta N^2(0)\rangle$, we need to assume a probability distribution
for $\delta N(0)$. Assuming Maxwell-Boltzmann Statistics, the probability distribution is 
given by, $N_c \exp{(\delta S)}$, where, $N_c$ is a normalisation
constant\cite{Landau}. Using \eqref{deltaS}, it can be written as $N_c
\exp{-\frac{1}{8}\frac{\delta A^2}{A_0}}$.  Switching to the variable
$\delta N$, this can be written as $N_c
\exp{\left( -\frac{\alpha^2}{2}{\delta N^2}{A_0}\right)}$. This implies
that\cite{Landau}
\begin{equation}
 \delta N^2(0)= \frac{A_0}{\alpha^2}. \label{DeltaN}
\end{equation}

From equations \eqref{zetaNR3}, \eqref{omega} and \eqref{DeltaN}; we get,\ 
\begin{equation}
 \zeta= -\frac{1}{16\pi},
\end{equation} 
exactly the value read off from the Navier-Stokes equation for the
horizon-fluid. It is to be noted here that the Principle of
Equipartition has been implicitly used here in writing down the
formula for $\zeta$ given by \eqref{zetaFa}.

\subsection{Positive Shear Viscosity Coefficient from the Theory of Fluctuations}

So far, we have shown that the anti-causal or teleological nature of the change in the area of the 
Horizon-fluid is responsible for the negative coefficient of Bulk Viscosity of the Horizon-fluid. 
Our analysis can not be extended to obtain the coefficient of Shear Viscosity. To get a better 
understanding, let us look at the evolution equations for the volume expansion coefficient 
and the shear of a null congruence along the event horizon of a Black Hole:
\begin{eqnarray}
\label{theta}
 -\frac{d\theta^H}{dt}+ g_H\theta^H &\simeq & 8\pi \mathcal{I}^H,  \\
\label{sigma}
 -\frac{d\sigma_{ab}^H}{dt}+ g_H\sigma_{ab}^H &=& \mathcal{E}_{ab}^H, 
\end{eqnarray}
In the above expressions, $g_H =2\pi T$, $\mathcal{I}^H$ denotes the energy flux through the horizon and $\mathcal{E}_{ab}^H$ 
is the Electric part of the Weyl tensor, i. e. the trace-free part of the Riemann tensor. 
For simplicity,  the quadratic term containing $\sigma$ in \eqref{theta} has been neglected.


It is important to note that although Eqs. \eqref{theta} and \eqref{sigma} look similar, actually they are 
quite different. This is because the source terms for \eqref{theta} and \eqref{sigma} are $\mathcal{I}^H$ and 
$\mathcal{E}_{ab}^H$ respectively and they are different in nature. 

While, $\mathcal{I}^H$ has its origins from the matter-field,  $\mathcal{E}_{ab}^H$ --- the
electric part of the Weyl tensor corresponding to the Gravitational field --- is related to the Gravitational 
waves\footnote{In case of vacuum solutions, the trace part of the Riemann tensor will vanish 
due to the Einstein equations, but the Weyl tensor can still be non-zero.}. For black-hole backgrounds,
such ingoing gravitational waves are the Quasinormal modes.  Quasinormal modes are the modes of energy 
dissipation and they decay with time. The decay of these modes automatically allows only the causal response. 
Hence, the coefficient of Shear Viscosity is positive. However, the driving term $\mathcal{I}^H$ for the 
volume expansion of  the congruence cannot be damped as it is not driven by Quasinormal modes. Hence, 
imposing a teleological boundary condition leads to negative bulk viscosity.  We note here that the ingoing 
boundary condition for the Gravitational waves at the Black Hole horizon is a physical one and is different 
from the boundary condition usually chosen in the Membrane paradigm. This difference has been noted in the 
literature in the context of the hydrodynamic limit of AdS-CFT\cite{MembrHolgra}.

\section{Discussion}
It is known for a long time that the event horizon of a black-hole behaves like a viscous fluid. This suggests that Gravity is emergent. 
However, the Horizon fluid is unusual --- it is a one parameter system and has negative
bulk viscosity.  There is no clear understanding of the statistical nature of
such a system in the literature. In this work, we have tried
to fill in that gap by constructing a theory of the fluctuations for
this system. Such a theory goes further than fluid mechanics and
provides a clear mathematical tool to evaluate certain parameters of the response functions whose values  
cannot be determined within fluid mechanics. It also establishes
the statistical mechanical nature of the horizon-fluid on a firmer
basis.

The analysis presented here is specific for Schwarzschild black-holes. However, 
the formalism developed here is sufficiently general to be applied to the 
fluid description for other black-holes. Also the fact that we use the Mean Field Theory 
to compute $\zeta$ does not make our analysis restrictive; it is model independent. 
Mean Field Theory fails to provide a good description of the 
horizon fluid when the Theory of Gravity is different from GR\cite{Louis}. 

The calculation of the Coefficient of bulk viscosity, $\zeta$ of the horizon-fluid from the
theory developed here leads to the exact value read off from the Navier-Stokes 
equation for the horizon-fluid. It also
provides a natural explanation for the negative value of $\zeta$ by
relating it with the teleological nature of black-hole event
horizons. Any Statistical Mechanical explanation for why the Bulk
Viscosity is negative has been lacking so far. This work addresses
that gap in the literature and provides insight into
this issue. An interesting corollary of our work is that the Coefficient of Bulk Viscosity 
for the fluid corresponding to a local horizon-like structure should be positive. One need not 
impose a future boundary condition in such a case 
\cite{LocalHF} and the Coefficient of Bulk Viscosity is positive, thus providing further
support for our claim. But the present work does not constitute a full explanation. The shear
equation for the null congruences on the horizon also has to be solved
subject to a teleological boundary condition. Hence it has to be
demonstrated that unlike $\zeta$, the Coefficient of Shear Viscosity
comes out to be positive in spite of this. Here we have provided a suggestion why 
$\eta_S$ is positive. It is also significant that for a class of fluids, it 
has been shown in the literature\cite{antCausTranspo}, that an anti-H Theorem
leads to negative transport coefficients.  (It shows that entropy of the 
system decreases with time.) For horizon-fluid being the opposite 
of a normal fluid in this respect, it is expected that a H-Theorem would lead to negative $\zeta$, whereas an 
anti-H Theorem would give rise to a positive value of $\zeta$ in this case.

Our analysis is the first step to take the fluid-gravity duality for
asymptotically flat black-holes to the next level. For AdS-background
black-holes, there already exists such a description via the CFT on
the boundary\cite{Son},\cite{BVAdS}. For general black-hole spacetimes, this is not
the case. However, the present formalism can only address the
transport properties of the horizon-fluid. It does not tell us whether
a corresponding description exists on the Gravity side at the level of the fluctuations. This is a question 
that lies outside the scope of the present formalism. 

Our formalism can be extended to compute the shear viscosity, 
electrical conductivity etc for the horizon-fluid. One can also look at other black-hole
spacetimes like Kerr, Reisner-Nandstorm and so on. It would 
be interesting to have a Statistical Mechanical understanding as to why Black Holes 
in the AdS-background are different. However, as noted in \cite{MFT}, the horizon-fluid 
in AdS-background has a richer phase diagram, that shows the existence of a first order 
phase transition and a tri-critical point. So the Theory of the Fluctuations might have 
to be developed in a quite independent fashion for AdS-horizon-fluids. Such a task 
lies outside the scope of this paper. It would also be
interesting to look at other theories of Gravity. Finally, it is well
known that the Transport coefficients diverge as one reaches the
critical point\cite{Kadanoff2}. Our analysis here does not address
this issue as we assume that we are sufficiently far from the critical
point to neglect this. The phase transition that occurs in the
Mean Field Theory of the horizon-fluid is such that only the ordered phase
corresponds to the black-hole\cite{MFT}.  This means that as we start
moving from the black-hole phase to some other phase, the Transport
coefficients would change from the values given for a Horizon fluid
and ultimately diverge as we come sufficiently close to the critical
point. We hope to address this issue elsewhere. 

\section*{Acknowledgments}

The work is supported by Max Planck-India Partner Group on Gravity and
Cosmology. SS is partially supported by Ramanujan Fellowship of DST,
India.

\begin{center}
\section*{Appendix}
\end{center}

\appendix

\section{Fluctuations of the Horizon Fluid}

\subsection{The Physical Process First Law for the horizon-fluid}
 
 The Physical process First Law for a black-hole Thermodynamics can be
 stated as follows: The mass of a black-hole increases as a result
 of mass-energy falling through the event horizon of the black-hole. 
 The area of the cross section of the event horizon also increases in this process. 
 Let us denote the temperature of the black-hole by $T$, the increase in mass by 
 $\delta M$ and the increase in the area of the horizon by $\delta A$. This law states that
 they are related by the following equation for a Schwarzschild black-hole~\cite{Wald,Jacobson,Parentani},
 \begin{equation}
  \delta M= \frac{1}{4}T\delta A. \label{process1st}
 \end{equation}
 In the Fluid picture, this corresponds to the following process: 
 Initially a fluid is in a state slightly away from equilibrium. It
 slowly moves towards the equilibrium state, which is the end state of
 this process. The change in the total energy of the fluid is related
 to the change in the volume(which is area because the horizon Fluid
 is $(2+1)$-D) by \eqref{process1st}.

 Using \eqref{deltaphi}, it can be shown from this that\cite{GravEssay}, $\delta E= T \delta S$. 
 This is the statement of the First Law of Thermodynamics for the fluid system.

\subsection{Non-equilibrium dynamics and Transport process}
For a system that is slightly away from equilibrium, Onsager's hypothesis\cite{Onsager} 
is applicable. Assuming $k$ does not change in Eq.~\eqref{eta}, we can describe
the evolution of the order parameter $\eta$ by the simplest form of
the Langevin equation,
\begin{equation}
 \ddot{\eta}= -\beta\dot{\eta}+ F(t), \label{BrE}
\end{equation}
where, $F(t)$ is the random term and $ \beta\dot{\eta}$ is the
anti-damping term(Since the bulk viscosity of the horizon-fluid is
negative.) due to the bulk viscosity of the fluid. Here we have taken
only the effect of bulk viscosity of the fluid as we take it to be
homogeneous.

Now we define a quantity, $x =\frac{\dot{N}}{N}=\frac{\dot{A}}{A}$ and
from \eqref{BrE}, it can be showed that\cite{GravEssay},
\begin{equation}
 \frac{d\langle x\rangle}{dt}= C \, T \, \langle x\rangle \, - 
\, \frac{1}{2}\langle x\rangle^2 \, + \, 8k\alpha\rho_d \, 
  \label{BrRC1x} 
\end{equation}
where, $\rho_d$ is the dissipated energy density and $C$ is a constant. 

Since, $x=\theta$, $A$, the area of the null congruence and $t$, the
affine parameter along it\cite{Damour}; \eqref{BrRC1x} is similar to
the Raychaudhuri equation sans the shear term,
\begin{equation}
 \frac{d\theta}{dt}= 2\pi T\theta
-\frac{1}{2}\theta^2-8\pi T_{\alpha\beta}\xi^\alpha\xi^\beta, \label{RC}
\end{equation}%
where, $C=-2\pi$ and 
\begin{equation}
  \rho_d=\frac{1}{8k\alpha}(8\pi T_{\alpha\beta}\xi^\alpha\xi^\beta
+\langle\dot{\eta}\rangle^2). \label{diisp}
\end{equation}
It is to be noted that except for the values of the coefficients, the
signatures have turned out to be the same in both these equations.

\section{The Computation of $\zeta$ in Frequency space}

$\zeta$ can also be determined by computing the autocorrelation
function for $F_{Th}$ in the frequency space.  The method we use here
is again that given by Kubo in \cite{Kubo}. In this case, we deduce
the formula for $\zeta$ from a generalised Langevin equation ab
initio. As we shall see, this makes the underlying assumptions
clear. We shall basically follow Kubo's approach\cite{Kubo} by
adopting it to our case.

We start with a generalised Langevin equation which gives the dynamics
of the transport process in which $A$ changes by $\delta
A(t)$. Following Kubo\cite{Kubo}, we assume that the process $\delta
A(t)$ is stationary at equilibrium. Also here we shall assume that the
autocorrelation functions depend only on the time difference. As
before, there is a thermodynamic force $F_{Th}$ acting on the
system. Then the Langevin equation can be written as
\begin{equation}
  \dot{\delta A(t)}= -\int_{-\infty}^t\gamma(t-t')\delta A(t')dt'+F_{Th}(t).\label{Langv1}
\end{equation}
Because of Hooke's law, $\bar{F}_{Th}(t)\propto \delta A(t)$. Then for
$F_{Th}(t)=0$, one has
\begin{equation}
\dot{\delta A(t)}= -\int_{-\infty}^t\gamma(t-t')\bar{F}_{Th}(t')dt'. \label{Langv2}
\end{equation}
Taking Fourier transform on both sides of \eqref{Langv2}, one gets,
\begin{equation}
\delta\dot{A}[\omega]= -\gamma[\omega]\bar{F}_{Th}[\omega]. \label{FTgamma}
\end{equation}
Since, $\delta\dot{A}[\omega]= -\delta A(t=0)+i\omega\delta
A[\omega]$, we have
\begin{equation}
  \delta A[\omega]= -(-\frac{i\gamma[\omega]}{\omega})\bar{F}_{Th}[\omega].\label{FTgamma1}
\end{equation}
For bulk viscosity, one has the relation\cite{Onsager, Zwanzig},
\begin{equation}
  \delta A[\omega]= -\zeta[\omega]\bar{F}_{Th}[\omega]A. \label{BVFT}
\end{equation}
The factor $A$ is present to signify that it is the area strain,
$\frac{\delta A}{A}$, that is related to the stress $\bar{F}_{Th}$ by
the coefficient of bulk viscosity.

Comparing \eqref{FTgamma1} with \eqref{BVFT}, one gets,
\begin{equation}
 \zeta[\omega]= -\frac{i\gamma[\omega]}{A\omega}.\label{zetagamma}
\end{equation}
The above relation would be useful in determining $\zeta$. Taking
Fourier transform of \eqref{Langv1}, we get,
\begin{equation}
 \delta A[\omega]= \frac{1}{i\omega+ \gamma[\omega]}F_{Th}[\omega].\label{LangevFT1}
\end{equation}

Let us now consider the Fourier transform of the autocorrelation
function for $\delta\dot{A}(t)$. Using the condition of stationarity,
we can write this as,
\begin{equation}
  F.T.[\langle\delta\dot{A}(t_0)\delta\dot{A}(t_0+t)]= i\omega \langle\delta A^2(t_0)\rangle - (i\omega)^2\int_0^\infty\exp{-i\omega t}\langle\delta A(t_0)\delta A(t_0+t)\rangle dt. \label{adotACf}
\end{equation}
Now one can determine $F.T.[\langle\delta A(t_0)\delta
A(t_0+t)\rangle]$ from \eqref{Langv1}, when $F_{Th}(t)=0$. In this
case, \eqref{Langv1} gives after taking Fourier transform on both
sides,
\begin{equation}
 \int_0^\infty\exp{-i\omega t}\langle\delta A(0)\delta A(t)\rangle dt= \frac{\langle\delta A^2(0)\rangle}{i\omega+\gamma[\omega]}. \label{AACf}
\end{equation}
Using \eqref{AACf} in \eqref{adotACf}, one gets,
\begin{equation}
  F.T.[\langle\delta\dot{A}(t_0)\delta\dot{A}(t_0+t)]= \frac{i\omega\gamma[\omega]}{i\omega+ \gamma[\omega]}\langle\delta A^2(t_0)\rangle.\label{AdotACf1}
\end{equation}
From the Langevin equation \eqref{Langv1}, one gets,
\begin{eqnarray}
  && F_{Th}(t_0)= \delta\dot{A}(t_0), \nonumber \\
  && F_{Th}(t_0+t)= \delta\dot{A}(t_0+t)+ \int_{t_0}^{t_0+t} \gamma(t_0+t-t') \delta A(t') dt'. \label{Ftht}
\end{eqnarray}
From \eqref{Ftht}, we can write down an expression for the
autocorrelation function for $F_{Th}$ as follows.
\begin{equation}
  \langle F_{Th}(t_0)F_{Th}(t_0+t)\rangle= \langle\delta\dot{A}(t_0)\delta\dot{A}(t_0+t)\rangle+ \int_{t_0}^{t_0+t} \gamma(t_0+t-t')\langle\delta\dot{A}(t_0)\delta A(t')\rangle dt'. \label{Ftht2}
\end{equation}
Taking the Fourier transform of \eqref{Ftht2} on both sides and using
\eqref{adotACf}, we get,
\begin{equation}
  F.T.[\langle F_{Th}(t_0)F_{Th}(t_0+t)\rangle]= \frac{i\omega\gamma[\omega]}{i\omega+ \gamma[\omega]}\langle\delta A^2(t_0)\rangle+\gamma[\omega]\int_0^\infty\langle\delta\dot{A}(t_0)\delta A(t_0+t)\rangle \exp{(-i\omega t)} dt. \label{FACFT1}
\end{equation}
After some algebra, one can write this as
\begin{equation}
 F.T.[\langle F_{Th}(t_0)F_{Th}(t_0+t)\rangle]= \gamma[\omega] \langle\delta A^2(t_0)\rangle. \label{FACFT2}
\end{equation}
Now we assume that the Principle of Equipartition holds and write
\eqref{FACFT2} as,
\begin{equation}
 \gamma[\omega]= Im[\frac{1}{KT} \int_0^\infty \langle F_{Th}(t_0)F_{Th}(t_0+t)\rangle\exp{(-i\omega t)} dt]. \label{gammaFAC}
\end{equation}
Using \eqref{zetagamma}, one can write,
\begin{equation}
 \zeta[\omega]= -\lim_{\omega\rightarrow\omega_S}Im[\frac{i}{AKT\omega}\int_0^\infty\langle F_{Th}(t_0)F_{Th}(t_0+t)\rangle\exp{(-i\omega t)}dt], \label{zetaF}
\end{equation}
where, $\omega_S$ is the frequency corresponding to the system size as
discussed during the computation of $\zeta$ in Real space.  At this point, we can
use \eqref{Fac} to get,
\begin{equation}
 \zeta[\omega]= -\lim_{\omega\rightarrow\omega_S}Im[\frac{i}{AKT\omega}P^2\int_{-\infty}^\infty\langle\delta A(t)\delta A(0)\rangle\theta(-t)\exp{(i(\bar{\omega}-\omega)t)} dt].\label{zetaFr1}
\end{equation}
\eqref{zetaFr1} follows from \eqref{zetaF} once we assume that $\delta
A(t)$ has a time reversed solution such that $\delta A(t)= \delta
A(-t)$.  Now as in the computation of $\zeta$ in Real space, one has
to assume that the fluctuating quantity obeys a damped wave equation
and take the infrared limit. However, in this case, if we assume,
$\delta A(t)= \delta A(0)\exp{[\pm i(\omega+i\epsilon)t]}$, then we
would get a Dirac delta function. To avoid this, we assume that
$\delta A(t)$ has a very small spread in the frequency. This allows us
to write
\begin{eqnarray}
  \delta A(t) &&= \delta A(0)\int_{\varDelta\omega}\exp{[i(\bar{\omega}+i\epsilon)t]}d\bar{\omega}; \ \ \ \ t>0,
  \nonumber \\ &&= \delta A(0)\int_{\varDelta\omega}\exp{[-i(\bar{\omega}+i\epsilon)t]}d\bar{\omega};\ \ t<0,         \label{deltaomega}
\end{eqnarray}
where, $\varDelta\omega$ is the spread in frequency of the fluctuating
quantity. We can take it to be as small as we want. It is also to be
noted that the condition, $\delta A(t)=\delta A(-t)$ is satisfied for
this choice. Using \eqref{deltaomega} in \eqref{zetaFr1}, we get,
\begin{equation}
  \zeta[\omega]=- \frac{P^2\langle\delta A^2(0)\rangle}{AKT\omega_S}. \label{zetaFr2}
\end{equation}
Proceeding in the same way as in the computation of $\zeta$ in Real
Space, we have finally,
\begin{equation}
 \zeta= - \frac{1}{16\pi},
\end{equation}
as earlier.

\end{document}